# Beyond Standard Quantum Chromodynamics[*][†]

Stanley J. Brodsky

*Stanford Linear Accelerator Center*
*Stanford University, Stanford, California 94309*

## Abstract

Despite the many empirical successes of QCD, there are a number of intriguing experimental anomalies that have been observed in heavy flavor hadroproduction, in measurements of azimuthal correlations in deep inelastic processes, and in measurements of spin correlations in hadronic reactions. Such phenomena point to color coherence and multiparton correlations in the hadron wavefunctions and physics beyond standard leading twist factorization. Two new high precision tests of QCD and the Standard Model are discussed: classical polarized photoabsorption sum rules, which are sensitive to anomalous couplings and composite structure, and commensurate scale relations, which relate physical observables to each other without scale or scheme ambiguity. The relationship of anomalous couplings to composite structure is also discussed.

## 1. Introduction

One of the most important achievements of high energy physics has been the development of quantum chromodynamics. The physical world of hadronic and nuclear interactions appears to be well-explained in terms of a minimal set of fundamental color-triplet quark fields and color-octet gluon gauge fields obeying exact local SU(3)-color symmetry. With only a few exceptions, such as charm hadroproduction and spin correlations, the theory has been validated by a vast array of experimental tests, particularly in high momentum transfer reactions where perturbative analyses are possible. Many types of novel QCD phenomena dependent on color coherence and asymptotic freedom have been observed, such as jet production, the strong logarithmic rise of the photon structure function at large $x_{bj}$, the rapid rise of the proton structure function at small $x_{bj}$, rapidity gaps, hard pomeron structure functions, and color transparency. Recent improvements in lattice gauge theory now provide a remarkably accurate description of the heavy quarkonium spectra as well as a precise determination of the QCD coupling[1].

It is plausible that there is physics at high energy beyond standard QCD, such as the existence of new fields with higher color representations, e.g., quixes $[6_C]$ and queights $[8_C]$; scalar gluons; the squarks and gluinos of supersymmetry; or the leptoquarks of grand unified theories. It is conceivable that the existing quarks or gauge fields are themselves composite at short distances, as in preon or technicolor models, leading to anomalous couplings and excited states of the existing quark or gluon fields. Later in this review I will discuss *classical polarized photoabsorption sum rules* as tests of anomalous couplings and composite structure in the standard model, and *commensurate scale relations*, which relate physical observables to each other to provide high precision tests of QCD without scale or scheme ambiguity.

---

[*] Work supported by the Department of Energy, contract DE–AC03–76SF00515.
[†] Talk presented at 4th International Conference on Physics Beyond the Standard Model, Lake Tahoe, California, December 13–18, 1994.

## 2. Novel Phenomena in QCD

Even without physics beyond the Standard Model, QCD itself predicts a novel spectrum of color-singlet bound states, such as the gluonia $(gg)$, $(ggg)$, hybrid states $(q\bar{q}g)$, and molecular analogs such as the H di-Lambda $(udsuds)$ and nuclear-bound quarkonium $(\overline{Q}Qqqq)$. In the nuclear domain, QCD predicts phenomena beyond standard nuclear physics, such as hidden-color configurations in light-nuclei, and the breakdown of traditional Glauber multiple scattering theory due to color coherence and color-filtering. At high density or high temperature, one anticipates new phases of QCD such as a quark-gluon plasma. In the following I will briefly review several examples of novel QCD phenomena:

*Color Transparency.* QCD predicts that fluctuations of a hadron wavefunction with a small color dipole moment can pass through nuclear matter without significant interaction[2,3]. For example, in the case of large momentum transfer exclusive reactions where only small-size valence Fock state configurations enter the hard scattering amplitude, both the initial and final state interactions of the hadron states become negligible. Evidence for diminished nuclear absorption in large angle quasielastic $pp$ scattering in nuclei was in fact reported by a BNL group[4], but the effect seemed to disappear anomalously at the highest beam energies. A new high precision experiment is now in progress. Evidence for QCD "color transparency" has now also been reported in high $Q^2$ $\rho$ leptoproduction for both nuclear coherent $\mu A \to \mu \rho A$ and incoherent $\mu A \to \mu \rho N(A-1)$ reactions by the E665 experiment at Fermilab[5], The recent NE18 measurement of quasielastic electron-proton scattering at SLAC finds results which do not clearly distinguish between conventional Glauber theory predictions and PQCD color transparency[6]. Conversely, Fock states with large-scale color configurations are predicted to strongly interact with high particle number production[7].

*Hidden Color.* The deuteron form factor at high $Q^2$ is sensitive to wavefunction configurations where all six quarks overlap within an impact separation $b^{\perp i} < \mathcal{O}(1/Q)$; the leading power-law falloff predicted by QCD[8] is $F_d(Q^2) = f(\alpha_s(Q^2))/(Q^2)^5$, where, asymptotically, $f(\alpha_s(Q^2)) \propto \alpha_s(Q^2)^{5+2\gamma}$. The derivation of the evolution equation for the deuteron distribution amplitude and its leading anomalous dimension $\gamma$ is given in Ref. 9. In general, the six-quark wavefunction of a deuteron is a mixture of five different color-singlet states. The dominant color configuration at large distances corresponds to the usual proton-neutron bound state. However at small impact space separation, all five Fock color-singlet components eventually acquire equal weight; *i.e.*, the deuteron wavefunction evolves to 80% "hidden color". The relatively large normalization of the deuteron form factor observed at large $Q^2$ points to sizeable hidden color contributions[10].

*Spin-Spin Correlations in Nucleon-Nucleon Scattering and the Charm Threshold.* One of the most striking anomalies in elastic proton-proton scattering is the large spin correlation $A_{NN}$ observed at large angles[11]. At $\sqrt{s} \simeq 5$ GeV, the rate for scattering with incident proton spins parallel and normal to the scattering plane is four times larger than scattering with antiparallel polarization. This strong polarization correlation can be attributed to the onset of charm production in the intermediate state at this energy[12]. The intermediate state $|uuduudc\bar{c}\rangle$ has odd intrinsic parity and couples to the $J = S = 1$ initial state, thus strongly enhancing scattering when the incident projectile and target protons have their spins parallel and normal to the scattering plane. The charm threshold can also explain the anomalous change in color transparency observed at the same energy in quasielastic $pp$ scattering. A crucial test is the observation of open charm production near threshold with a cross section of order of $1\mu$b.

*Anomalous Decays of the $J/\psi$.* The dominant two-body hadronic decay channel of the $J/\psi$ is $J/\psi \to \rho\pi$ even though such vector-psuedoscalar final states are forbidden in leading order by helicity conservation in perturbative QCD[13]. The $\psi'$, on the other hand, appears to respect PQCD. The $J/\psi$ anomaly may signal mixing with vector gluonia or other exotica.[13]

*The QCD Van Der Waals Potential and Nuclear Bound Quarkonium.* The simplest form of the nuclear force is the interaction between two heavy quarkonium states, such as the $\Upsilon(b\bar{b})$ and the $J/\psi(c\bar{c})$. Since there are no valence quarks in common, the dominant color-singlet interaction arises simply from the exchange of two or more gluons. In principle, one could measure the interactions of such systems by producing pairs of quarkonia in high energy hadron collisions. The same fundamental QCD van der Waals potential also dominates the interactions of heavy quarkonia with ordinary hadrons and nuclei. As shown in Ref. 14, the small size of the $Q\bar{Q}$ bound state relative to the much larger hadron sizes allows a systematic expansion of the gluonic potential using the operator product potential. The coupling of the scalar part of the interaction to large-size hadrons is rigorously normalized to the mass of the state via the trace anomaly. This scalar attractive potential dominates the interactions at low relative velocity. In this way one establishes that the nuclear force between heavy quarkonia and ordinary nuclei is attractive and sufficiently strong to produce nuclear-bound quarkonium[14,15].

*Leading Particle Effect in Open Charm Production.* According to PQCD factorization, the fragmentation of a heavy quark jet is independent of the production process. However strong correlations between the quantum numbers of $D$ mesons and the charge of the incident pion beam in $\pi N \to DX$ reactions. This effect can be explained as due to the coalescence of the produced charm quark with co-moving valence quarks. The same higher-twist recombination effect can also account for the suppression of $J/\psi$ and $\Upsilon$ production in nuclear collisions in phase space regions of high particle density.[16]

*Anomalous Quarkonium Production at the Tevatron.* Strong discrepancies between conventional QCD predictions and experiment of a factor of 30 or more have recently been observed for $\psi$, $\psi'$, and $\Upsilon$ production at large $p_T$ in high energy $p\overline{p}$ collisions at the Tevatron[17]. Braaten and Fleming[18] have suggested that the surplus of charmonium production is due to the enhanced fragmentation of gluon jets coupling to the octet $c\overline{c}$ components in higher Fock states $|c\overline{c}gg\rangle$ of the charmonium wavefunction. Such Fock states are required for a consistent treatment of the radiative corrections to the hadronic decay of P-waves in QCD[19]. However, it is not clear whether this proposal can also solve the large discrepancies observed in $\Upsilon$ production. Also, as I shall review in the next section there are many other anomalies observed in charm hadroproduction which are incompatible with standard leading twist PQCD factorization.

## 3. Higher Twist Contributions in QCD

Higher twist corrections are an inevitable complication in QCD predictions. Power-suppressed corrections arise from non-perturbative corrections to the gluon and quark propagators, mass insertions, etc. One also expects dynamical higher twist contributions involving more than one parton in the hadron wavefunction. For example, at large values of the quarkonium momentum fraction $x_F$, it becomes advantageous for two or more collinear partons from the projectile to participate in the reaction. Such processes are suppressed relative to ordinary fusion reactions by powers of $\Lambda_{QCD}/m_Q$ where $\Lambda_{QCD}$ is the characteristic transverse momentum in the incident hadron wavefunction. Despite the extra powers of $1/m_Q$, the multiparton processes can become dominant at $(1-x_F) < \mathcal{O}(\Lambda_{QCD}^2/m_Q^2)$ since they are efficient in converting the incident hadron momentum into high $x_F$ quarkonia [20]. Similarly, in deep inelastic lepton-proton scattering, there are higher twist contributions from the interference of amplitudes where the lepton hits different quarks.

*Higher Twist Contributions in the Drell-Yan Process.* In the $x_F \to 1$ limit, important higher twist effects are expected [21,22] and observed [23] in the muon pair production process, $\pi N \to \mu^+\mu^- + X$. In effect, both valence quarks in the pion projectile must be involved in the reaction if the full momentum is to be delivered to the muons. The higher twist effect manifests itself in the angular distribution of the muons: the polarization of the virtual photon changes from transverse to longitudinal at large $x_F$. Thus the photon tends to carry the same helicity as the pion in the $x_F \to 1$ limit. Recently, Brandenburg, Khoze, Müeller, and I[24] have shown that the same higher twist mechanism also accounts for the anomalously large $cos\phi$ and $cos2\phi$ azimuthal correlations observed in the Drell-Yan process. The size of these correlations also places constraints on the shape of the projectile distribution amplitude.

*Evidence for Higher Twist Contributions in Quarkonium Production.* Quarkonium bound states formed by heavy quark-antiquark pairs are small nonrelativistic systems, whose production and decay properties are expected to be governed by perturbative QCD. In leading twist QCD the production of the $J/\psi$ at low transverse momentum occurs both "directly" from the gluon fusion subprocess $gg \to J/\psi + g$ and indirectly via the production of $\chi_1$ and $\chi_2$ states. At high transverse momentum, one also has to take into account production through quark and gluon fragmentation [25]. Recent E705 and E672 data [26,27] on the production fractions of the various charmonium states have confirmed that there is a clear discrepancy with the leading twist QCD prediction. The recent leading-twist analysis of Vantinnen, et al.[28] shows that the predicted ratio of direct $J/\psi$ production in $\pi N$ collisions compared to the $\chi_2$ production is too low by a factor of about 3. In addition, the ratio of $\chi_1$ production to $\chi_2$ production is too low by a factor of 10. A similar conclusion has been reached in Ref. 29, where possible explanations in terms of uncertainties in the partonic cross sections (very different $K$-factors for the various processes) or unconventional pion parton distributions are discussed.

The wealth of data from the NA3 experiment at CERN[30] and the Chicago-Iowa-Princeton[31] and E537 experiments[32] at FermiLab on the angular distribution of the muons in the decay $J/\psi \to \mu^+\mu^-$ provides an even more sensitive discriminant of different production mechanisms [33-39]. The polarization of the $c\bar{c}$, and hence that of the charmonium bound state[34], can at leading twist be calculated from perturbative QCD. Furthermore, in the heavy quark limit, the radiative transition $\chi_J \to J/\psi + \gamma$ preserves the quark spins, i.e., it is an electric dipole transition. Hence the polarization also of indirectly produced $J/\psi$'s can be calculated. Even if the relative production rates of the $J/\psi$, $\chi_1$ and $\chi_2$ are adjusted (using $K$-factors) to agree with the data, the $J/\psi$ polarization data is still not reproduced[28]. The direct $J/\psi$ and $\chi_1$ subprocesses require, at leading order and twist, the emission of a quark or gluon, e.g., $gg \to J/\psi + g$. This implies a higher subenergy $\sqrt{\hat{s}}$ for these processes compared to that for the $\chi_2$, which can be produced through simple gluon fusion, $gg \to \chi_2$. It is then plausible that a higher twist component which avoids the necessity for gluon emission is more significant for the $J/\psi$ and the $\chi_1$ than it is for the $\chi_2$.

It is thus natural to expect dynamical higher twist effects to be enhanced in $J/\psi$ production at large $x_F$. The data does indeed show a remarkable turnover in the polarization of the $J/\psi$ for $x_F \gtrsim 0.8$, with the fastest $J/\psi$'s being longitudinally polarized. Additional independent evidence for higher twist effects in $J/\psi$ production is reflected in the nuclear target $A$-dependence of the cross section. In lepton pair production, the cross section is very closely linearly dependent on $A$ (apart from a small deviation at the largest $x_F$ [40]). $J/\psi$ production, on the other hand, shows a nuclear suppression over the whole $x_F$ range [41]. The suppression depends on $x_F$ rather than on $x_2$[42]. QCD factorization is thus broken, implying that the effect is due to higher twist terms.

*Intrinsic Heavy Quark Contributions in Hadron Wavefunctions.* The QCD wavefunction of a hadron can be represented as a superposition of quark and gluon Fock states. For example, at fixed light-cone time, $\tau = t + z/c$, the $\pi^-$ wavefunction can be expanded as a sum over the complete basis of free quark and gluon states: $|\Psi_{\pi^-}\rangle = \sum_n |n\rangle \psi_{n/\pi^-}(x_i, k_{T,i}, \lambda_i)$ where the color-singlet states, $|n\rangle$, represent the Fock components $|\bar{u}d\rangle$, $|\bar{u}dg\rangle$, $|\bar{u}dQ\bar{Q}\rangle$, etc. Microscopically, the intrinsic heavy quark Fock component in the $\pi^-$ wavefunction, $|\overline{u}dQ\overline{Q}\rangle$, is generated by virtual interactions such as $gg \to Q\overline{Q}$ where the gluons couple to two or more projectile valence quarks. The probability for $Q\overline{Q}$ fluctuations to exist in a light hadron thus scales as $\alpha_s^2(m_Q^2)/m_Q^2$ relative to leading-twist production[16]. Therefore, this contribution is higher twist, power-law suppressed compared to sea quark contributions generated by gluon splitting. When the projectile scatters in the target, the coherence of the Fock components is broken, its fluctuations can hadronize, forming new hadronic systems from the fluctuations[20]. For example, intrinsic $c\overline{c}$ fluctuations can be liberated provided the system is probed during the characteristic time, $\Delta t = 2p_{\rm lab}/M_{c\overline{c}}^2$, that such fluctuations exist. For soft interactions at momentum scale $\mu$, the intrinsic heavy quark cross section is suppressed by an additional resolving factor $\propto \mu^2/m_Q^2$ [43]. The nuclear dependence arising from the manifestation of intrinsic charm is expected to be $\sigma_A \approx \sigma_N A^{2/3}$, characteristic of soft interactions.

In general, the dominant Fock state configurations are not far off shell and thus have minimal invariant mass, $M^2 = \sum_i m_{T,i}^2/x_i$ where $m_{T,i}$ is the transverse mass of the $i^{\rm th}$ particle in the configuration. Intrinsic $Q\overline{Q}$ Fock components with minimum invariant mass correspond to configurations with equal rapidity constituents. Thus, unlike sea quarks generated from a single parton, intrinsic heavy quarks tend to carry a larger fraction of the parent momentum than the light quarks[44]. In fact, if the intrinsic $Q\overline{Q}$ coalesces into a quarkonium state, the momentum of the two heavy quarks is combined so that the quarkonium state will carry a significant fraction of the projectile momentum.

There is substantial evidence for the existence of intrinsic $c\overline{c}$ fluctuations in the wavefunction of light hadrons. For example, the charm structure function of the proton measured by EMC is significantly larger than predicted by photon-gluon fusion at large $x_{Bj}$ [45]. Leading charm production in $\pi N$ and hyperon-$N$ collisions also requires a charm source beyond leading twist [16,46]. The NA3 experiment has also shown that the single $J/\psi$ cross section at large $x_F$ is greater than expected from $gg$ and $q\overline{q}$ production[47]. The nuclear dependence of this forward component is diffractive-like, as expected from the BHMT mechanism. Also, as we have noted above, intrinsic charm may account for the anomalous longitudinal polarization of the $J/\psi$ at large $x_F$[48] seen in $\pi N \to J/\psi X$ interactions.

Further theoretical work is needed to establish that the data on direct $J/\psi$ and $\chi_1$ production indeed can be described using a higher twist intrinsic charm mechanism as discussed in Ref. 20. Experimentally, it is important to check whether the $J/\psi$'s produced indirectly via $\chi_2$ decay are transversely polarized. This would show that $\chi_2$ production is dominantly leading twist, as we have argued. Better data on real or virtual photoproduction of the individual charmonium states would also add important information.

*Double Quarkonium Hadroproduction.* It is quite rare for two charmonium states to be produced in the same hadronic collision. However, the NA3 collaboration has measured a double $J/\psi$ production rate significantly above background in multi-muon events with $\pi^-$ beams at laboratory momentum 150 and 280 GeV/c[49] and a 400 GeV/c proton beam[50]. The relative double to single rate, $\sigma_{\psi\psi}/\sigma_\psi$, is $(3\pm1)\times10^{-4}$ for pion-induced production where $\sigma_\psi$ is the integrated single $\psi$ production cross section. A particularly surprising feature of the NA3 $\pi^-N \to \psi\psi X$ events is that the laboratory fraction of the projectile momentum carried by the $\psi\psi$ pair is always very large, $x_{\psi\psi} \geq 0.6$ at 150 GeV/c and $x_{\psi\psi} \geq 0.4$ at 280 GeV/c. In some events, nearly all of the projectile momentum is carried by the $\psi\psi$ system! In contrast, perturbative $gg$ and $q\bar{q}$ fusion processes are expected to produce central $\psi\psi$ pairs, centered around the mean value, $\langle x_{\psi\psi}\rangle \approx 0.4\text{-}0.5$, in the laboratory. There have been attempts to explain the NA3 data within conventional leading-twist QCD. Charmonium pairs can be produced by a variety of QCD processes including $B\overline{B}$ production and decay, $B\overline{B} \to \psi\psi X$ [51] and $\mathcal{O}(\alpha_s^4)\,\psi\psi$ production via $gg$ fusion and $q\bar{q}$ annihilation[52–54]. Li and Liu have also considered the possibility that a $2^{++}c\bar{c}c\bar{c}$ resonance is produced which then decays into correlated $\psi\psi$ pairs[55]. All of these models predict centrally produced $\psi\psi$ pairs[51–54], in contradiction to the $\pi^-$ data. In addition, the predicted magnitude of $\sigma_{\psi\psi}$ is too small by a factor of 3-5. If these models are updated using recent branching ratios and current scale-dependent parton distributions, the predicted leading twist cross sections are further reduced, suggesting that an additional mechanism is needed to produce fast $\psi\psi$ pairs.

Over a sufficiently short time, the pion can contain Fock states of arbitrary complexity. For example, two intrinsic $c\bar{c}$ pairs may appear simultaneously in the quantum fluctuations of the projectile wavefunction and then, freed in an energetic interaction, coalesce to form a pair of $\psi$'s. Ramona Vogt and I have recently made a model calculation of double charmonium production based on a light-cone Fock state wavefunction which is approximately constant up to the energy denominator. The predicted $\psi\psi$ pair distributions from the intrinsic charm model provides a natural explanation of the strong forward production of double $J/\psi$ hadroproduction and thus gives strong phenomenological support for the presence of intrinsic heavy quark states in hadrons[56].

It is clearly important for the double $J/\psi$ measurements to be repeated with higher statistics and also at higher energies. The same intrinsic Fock states will also lead to the production of multi-charmed baryons in the proton fragmentation region. The intrinsic heavy quark model can also be used to predict the features of heavier quarkonium hadroproduction, such as $\Upsilon\Upsilon$, $\Upsilon\psi$, and $(c\bar{b})\,(\bar{c}b)$ pairs. It is also interesting to study the correlations of the heavy quarkonium pairs to search for possible new four-quark bound states and final state interactions generated by multiple gluon exchange[55] since the QCD Van der Waals interactions could be anomalously strong at low relative rapidity[14,15].

There are many ways in which the intrinsic heavy quark content of light hadrons can be tested. More measurements of the charm and bottom structure functions at large $x_F$ are needed to confirm the EMC data[45]. Charm production in the proton fragmentation region in deep inelastic lepton-proton scattering is sensitive to the hidden charm in the proton wavefunction. The presence of intrinsic heavy quarks in the hadron wavefunction also enhances heavy flavor production in hadronic interactions near threshold. More generally, the intrinsic heavy quark model leads to enhanced open and hidden heavy quark production and leading particle correlations at high $x_F$ in hadron collisions with a distinctive strongly-shadowed nuclear dependence characteristic of soft hadronic collisions.

## 4. Electromagnetic and Axial Moments of Relativistic Bound States

The magnetic moment of a non-relativistic bound state system can be computed simply by summing the moments of its constituents. The situation is much more interesting and complex for composite systems where relativistic recoil effects must be taken into account. For example, at infinitely small radius $R_p M_p \to 0$, the magnetic moment of a proton must become equal to the Dirac moment $e/2M_p$, as demanded by the Drell-Hearn-Gerasimov sum rule[57,58]. Similarly, in the case of spin-1 systems, the quadrupole moment becomes identical to $-e/M^2$ in the point-like limit[59]. Thus the deuteron quadrupole moment is in general nonzero even if the nucleon-nucleon bound state has no D-wave component[59]. Such effects are due to the fact that even "static" moments have to be computed as transitions between states of different momentum $p^\mu$ and $p^\mu + q^\mu$ with $q^\mu \to 0$. Thus one must construct current matrix elements between boosted states. The Wigner boost generates nontrivial corrections to the current interactions of bound systems[60], and in the point-like limit, these generate the canonical couplings.

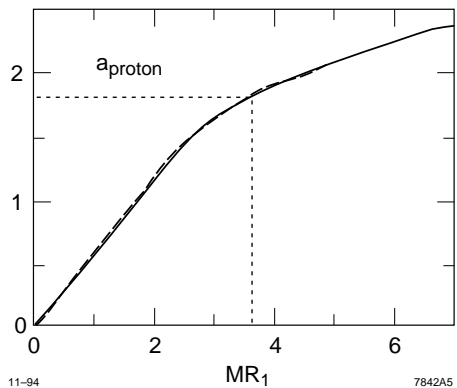

Figure 1. The anomalous magnetic moment $a = F_2(0)$ of the proton as a function of $M_p R_1$: broken line, pole type wavefunction; continuous line, gaussian wavefunction. The experimental value is given by the dotted lines. The prediction of the model is independent of the wavefunction for $Q^2 = 0$.

Felix Schlumpf and I have recently used a three-quark light-cone model to display the functional relationship between the anomalous moment $a_p$ and its Dirac radius[61]. The result is shown in Fig. 1. The value of $R_1^2 = -6dF_1(Q^2)/dQ^2|_{Q^2=0}$ is varied by changing the size parameters in the Figure 1 shows that when one plots the dimensionless observable $a_p$ against the dimensionless observable $MR_1$ the prediction is essentially independent of the assumed power-law or Gaussian form of the three-quark light-cone wavefunction. The only parameter controlling the relation between the dimensionless observables in the light-cone three-quark model is $m/M_p$ which is set to 0.28. For the physical proton radius $M_p R_1 = 3.63$ one obtains the empirical value for $a_p = 1.79$ (indicated by the dotted lines in Fig. 1). The same three-quark model also gives $g_A = 1.25$ for the non-singlet axial coupling in agreement with experiment The singlet helicity sum $\Delta\Sigma$ for the three quark model is predicted to be 0.75. This will be substantially reduced when gluon and sea quark Fock state contributions are included.

The light-cone model predicts that the quark helicity sum $\Delta\Sigma = \Delta u + \Delta d$ and $g_A = \Delta u - \Delta d$ vanishes as a function of the proton radius $R_1$ in a similar way as the anomalous moment vanishes. Since the helicity sum $\Delta\Sigma$ depends on the proton size, it clearly cannot be identified as the vector sum of the rest-frame constituent spins. Note that $\Delta q$ refers to the difference of helicities at fixed light-cone time or at infinite momentum; it cannot be identified with $q(s_z = +\frac{1}{2}) - q(s_z = -\frac{1}{2})$, the spin carried by each quark flavor in the proton rest frame in the equal time formalism[62,61]. In fact, $\Delta q$ vanishes as $R_1 \to 0$ since the constituent quark helicities become completely disoriented for large internal transverse momentum.

The above results have important implications for theories in which leptons, quarks, or gauge particles are composite at short distances. If the internal scale of such a theory is sufficiently high, then the Drell-Hearn Gerasimov (DHG) sum rule[57] guarantees that the magnetic and quadrupole couplings of the composite states are indistinguishable from those of the Standard Model. However, in the conventional light-cone bound state formalism, a high internal momentum scale drives the axial coupling of the composite system to zero rather than the standard canonical coupling.

## 5. Classical Polarized Photoabsorption Sum Rules

The Dirac value $g = 2$ for the magnetic moment $\mu = geS/2M$ of a particle of charge $e$, mass $M$, and spin $S$, plays a special role in quantum field theory. As shown by Weinberg[63] and Ferrara et al.[64], the canonical value $g = 2$ gives an effective Lagrangian which has maximally convergent high energy behavior for fields of any spin. In the case of the Standard Model, the anomalous magnetic moments $\mu_a = (g-2)eS/2M$ and anomalous quadrupole moments $Q_a = Q + e/M^2$ of the fundamental fields vanish at tree level, ensuring a quantum field theory which is perturbatively renormalizable. However, as discussed in the previous section, one can use the DHG sum rule[57] to show that the magnetic and quadrupole moments of spin-$\frac{1}{2}$ or spin-1 bound states approach the canonical values $\mu = eS/M$ and $Q = -e/M^2$ in the zero radius limit $MR \to 0$[58,61,59], independent of the internal dynamics. Deviations from the predicted values will thus reflect new physics and interactions such as virtual corrections from supersymmetry or an underlying composite structure.

The canonical values $g = 2$ and $Q = -e/M^2$ lead to a number of important phenomenological consequences: (1) The magnetic moment of a particle with $g = 2$ processes with the same frequency as the Larmor frequency in a constant magnetic field. This synchronicity is a consequence of the fact that the electromagnetic spin currents can be formally generated by an infinitesimal Lorentz transformation[65,66]. (2) The forward helicity-flip Compton amplitude for a target with $g = 2$ vanishes at zero energy[67]. (3) The Born amplitude for a photon radiated in the scattering of any number of incoming and outgoing particles with charge $e_i$ and four-momentum $p_i^\mu$ vanishes at the kinematic angle where all the ratios $e_i/p_i \cdot k$ are simultaneously equal[66]. For example, the Born cross section $d\sigma/\cos\theta_{cm}(u\overline{d} \to W^+\gamma)$ vanishes identically at an angle determined from the ratio of charges: $\cos\theta_{cm} = e_d/e_{W^+} = -1/3$[68]. Such "radiative amplitude zeroes" or "null zones" occur at lowest order in the Standard Model because the electromagnetic spin currents of the quarks and the vector gauge bosons are all canonical.

The vanishing of the forward helicity-flip Compton amplitude at zero energy for the canonical couplings, together with the optical theorem and dispersion theory, leads to a superconvergent sum rule; *i.e.*, a zero value for the DHG sum rule. This remarkable observation was first made for quantum electrodynamics and the electroweak theory by Altarelli, Cabibbo and Maiani[69]. Recently, Ivan Schmidt and I[70] have used a quantum loop expansion to show that the logarithmic integral of the spin-dependent part of the photoabsorption cross section

$$\int\limits_{\nu_{th}}^{\infty} \frac{d\nu}{\nu} \Delta\sigma_{\text{Born}}(\nu) = 0 \tag{1}$$

for any $2 \to 2$ Standard Model process $\gamma a \to bc$ in the classical, tree graph approximation. The particles $a, b, c$ and $d$ can be leptons, photons, gluons, quarks, elementary Higgs particles, supersymmetric particles, etc. We also can extend the sum rule to certain virtual photon processes. Here $\nu = p \cdot q/M$ is the laboratory energy and $\Delta\sigma(\nu) = \sigma_P(\nu) - \sigma_A(\nu)$ is the difference between the photoabsorption cross section for parallel and antiparallel photon and target helicities. The sum rule receives nonzero contributions in higher order perturbation theory in the Standard Model from both quantum loop corrections and higher particle number final states. Similar arguments also imply that the DHG integral vanishes for virtual photoabsorption processes such as $\ell\gamma \to \ell Q\overline{Q}$ and $\ell g \to \ell Q\overline{Q}$, the lowest order sea-quark contribution to polarized deep inelastic photon and hadron structure functions. Note that the integral extends to $\nu = \nu_{th}$, which is generally beyond the usual leading twist domain.

We can use Eq. (1) as a new way to test the canonical couplings of the Standard Model and to isolate the higher order radiative corrections. The sum rule also provides a non-trivial consistency check on calculations of the polarized cross sections. Probably the most interesting application and test of the Standard

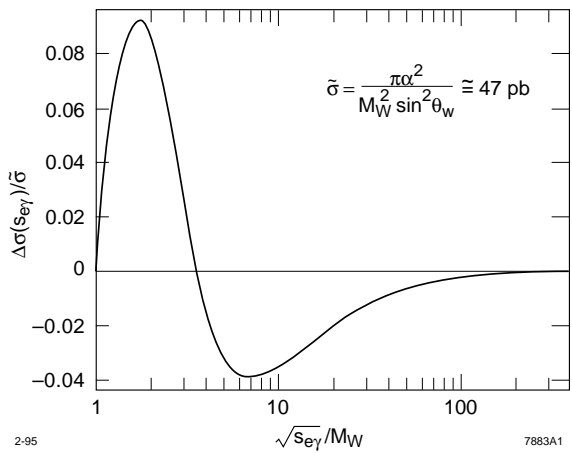

Figure 2. The Born cross section difference $\Delta\sigma$ for the Standard Model process $\gamma e \to W\nu$ for parallel minus antiparallel electron/photon helicities as a function of $\log \sqrt{s}_{e\gamma}/M_W$ The logarithmic integral of $\Delta\sigma$ vanishes in the classical limit.

Model is to the reactions $\gamma\gamma \to q\bar{q}$, $\gamma e \to W\nu$ and $\gamma e \to Ze$ which can be studied in high energy polarized electron-positron colliders with back-scattered laser beams. In contrast to the timelike process $e^+e^- \to W^+W^-$, the $\gamma\gamma$ and $\gamma e$ reactions are sensitive to the anomalous moments of the gauge bosons at $q^2 = 0$. The cancellation of the positive and negative contributions[71] of $\Delta\sigma(\gamma e \to W\nu)$ to the DHG integral is evident in Fig. 2.

The vanishing of the logarithmic integral of $\Delta\sigma(\nu)$ at the tree-graph approximation also implies that there must be an energy $\nu_0$ where $\Delta\sigma_{\text{Born}}(\nu_0) = 0$[72]. Modifications of the Standard Model, such as those arising from composite structure of the quarks or vector bosons, will lead to corrections to the sum rule. Tom Rizzo, Ivan Schmidt, and I[72] have investigated the sensitivity of the position of the crossing point $\sqrt{s}_{\gamma e} \simeq 3.16 M_W$ and the value of the DHG integral to higher order corrections and violations of the Standard Model. These results can clearly be generalized to other higher order tree-graph processes in the Standard Model and supersymmetric gauge theory.

# 6. Commensurate Scale Relations: Precise Tests of Quantum Chromodynamics Without Scale or Scheme Ambiguity

The renormalization scale dependence of perturbative QCD predictions has plagued attempts to make high precision tests of the theory. The problem is compounded in multi-scale problems where several plausible physical scales enter. Recently Hung Jung Lu and I have shown how the scale ambiguity problem can be avoided by focussing on relations between experimentally-measurable observables[73]. For example, consider the entire radiative corrections to the annihilation cross section expressed as the "effective charge" $\alpha_R(Q)$ where $Q = \sqrt{s}$: $R(Q) \equiv 3 \sum_f Q_f^2 \left[1 + \alpha_R(Q)/\pi\right]$. Similarly, we can define the entire radiative correction to the Bjorken sum rule as the effective charge $\alpha_{g_1}(Q)$ where $Q$ is the lepton momentum transfer: $\int_0^1 dx \left[g_1^{ep}(x,Q^2) - g_1^{en}(x,Q^2)\right] \equiv (1/3)(g_A/g_V)\left[1 - \alpha_{g_1}(Q)/\pi\right]$. We now use the known expressions to three loops in $\overline{MS}$ scheme and choose the scales $Q^*$ and $Q^{**}$ as in the BLM method[74] to re-sum all non-conformal contributions from the QCD $\beta$–function into the running couplings. This prescription ensures that, as in quantum electrodynamics, all vacuum polarization contributions are incorporated into the coupling rather than the coefficients. The values of these scales are the physical values of the energies or momentum transfers which ensure that the radiative corrections to each observable passes through the heavy quark thresholds at their respective commensurate physical scales. The result is remarkably simple:

$$\frac{\alpha_{g_1}(Q)}{\pi} = \frac{\alpha_R(Q^*)}{\pi} - \left(\frac{\alpha_R(Q^{**})}{\pi}\right)^2 + \left(\frac{\alpha_R(Q^{***})}{\pi}\right)^3 + \cdots. \qquad (2)$$

It is remarkable that the coefficients in Eq. (2) form a geometric series. In fact in the conformal limit where $\beta = 0$ and $\alpha$ is constant we recover the Crewther relation[75,76] $(1 + \alpha_R/\pi)(1 - \alpha_{g_1}/\pi) = 1$. Thus Eq. (2) can be regarded as the extension of the Crewther relation to non-conformally invariant gauge theory.

Hung Jung Lu and I refer to the connections between the effective charges of observables such as Eq. (2)) as "commensurate scale relations" (CSR)[73]. QCD observables must track in both normalization and shape as given by the CSR. Although the conventional $\overline{MS}$ scheme is used as an intermediary, the final relations between observables are independent of theoretical conventions such as the choice of intermediate renormalization scheme and scale as is required by renormalization group invariance[77]. The commensurate scale relations thus provide fundamental tests of QCD which can be made increasingly precise without scale or scheme ambiguity. Since the ambiguities due to scale and scheme choice have been eliminated, one can ask fundamental questions concerning the influence of higher twist terms, the nature of the QCD perturbative expansions, *e.g.*, whether the series is convergent or asymptotic, due to renormalons, etc.[78,79]

A natural procedure for developing a precision QCD phenomenology will be to choose one effective charge as the canonical definition of the QCD coupling, and then predict all other observables in terms of this canonical measure. Ideally, the heavy quark effective charge $\alpha_V(Q^2)$ could serve this central role since it can be determined from both the quarkonium spectrum and from lattice gauge theory. There is an intrinsic disadvantage in using $\alpha_{\overline{MS}}(Q)$ as an expansion parameter: the function $\alpha_{\overline{MS}}(Q)$ has a simple pole at $Q = \Lambda_{\overline{MS}}$, whereas observables are by definition finite.

A number of examples of three-loop commensurate scale relations are given in Ref. 73. The BLM method has also been applied to the analysis of jet ratios in $ep$ collisions by Ingelman and Rathsman [80]. One can determine the scale $Q^*$ for $(2+1)$ jets at HERA as a function of all of the available scales. In the case of jet production at the Z, Kramer and Lampe[81] find that the BLM scale and the NLO PQCD predictions give a consistent description of the LEP 2-jet and 3-jet data with a value for $\Lambda_{\overline{MS}}$ considerably smaller than conventional analyses. It is clear that a comprehensive reanalysis of the SLD and LEP data is needed.

The BLM method and commensurate scale relations can be applied to the whole range of QCD and standard model processes, making the tests of theory much more sensitive. Recent applications include the radiative corrections to the top width decay by Voloshin and Smith[82] and to other electroweak measures by Sirlin[83]. One of the most interesting and important areas of application of commensurate scale relations will be to the hadronic corrections to exclusive and inclusive weak decays of heavy quark systems, since the scale ambiguity in the QCD radiative corrections is at present often the largest component in the theoretical error entering electroweak phenomenology.

### 6.1 ACKNOWLEDGMENTS

I thank Jack Gunion and his colleagues for organizing this interesting conference. Much of this work reported here was done in collaboration with others, including Paul Hoyer, Hung Jung Lu, Tom Rizzo, Felix Schlumpf, Ivan Schmidt, Wai-Keung Tang, Mikko Vänttinen, and Ramona Vogt. This work is supported in part by the Department of Energy, contract DE–AC03–76SF00515.

# REFERENCES

1. W. Dimm, G. P. Lepage, and P. B. Mackenzie, HEPLAT-9412100 (1994) and references therein.
2. G. Bertsch, S. J. Brodsky, A. S. Goldhaber, and J.F. Gunion, *Phys. Rev. Lett.* **47**, 297 (1981).
3. S. J. Brodsky and A. H. Mueller, *Phys. Lett.* **206B**, 685 (1988).
4. S. Heppelmann, *Nucl. Phys. B, Proc. Suppl.* **12**, 159 (1990), and references therein.
5. G. Fang, *et al.*, presented at the INT - Fermilab Workshop on *Perspectives of High Energy Strong Interaction Physics at Hadron Facilities* (1993). M. R. Adams, *et al.*, FERMILAB-PUB-94-233-E (1994). A. Kotwal, *et al.*, FERMILAB-Conf-94/345-E (1994).
6. N. Makins, *et al.*, (1994), NE-18 Collaboration. MIT preprint.
7. B. Blaettel, G. Baym, L. L. Frankfurt, H. Heiselberg, M. Strikman, *Phys. Rev.* **D47**, 2761 (1993).
8. S. J. Brodsky and B. T. Chertok, *Phys. Rev.* **D14**, 3003 (1976).
9. S. J. Brodsky, C.-R. Ji, and G. P. Lepage, *Phys. Rev. Lett.* **51**, 83 (1983).
10. G. R. Farrar, K. Huleihel, and H. Zhang, *Phys. Rev. Lett.* **74**, 650. (1995)
11. A. D. Krisch, *Nucl. Phys. B (Proc. Suppl.)* **25B**, 285. (1992)
12. S. J. Brodsky and G. F. de Teramond, *Phys. Rev. Lett.* **60**, 1924 (1988).
13. S. J. Brodsky, G. P. Lepage, and S. F.Tuan *Phys. Rev. Lett.* **59**, 621 (1987) and references therein.
14. M. Luke, A. V. Manohar, and M. J. Savage, *Phys. Lett.* **B288**, 355. (1992).
15. S. J. Brodsky, G. F. de Teramond, and I. A. Schmidt (1990), *Phys. Rev. Lett.* **64**, 1011 (1990).
16. R. Vogt, S. J. Brodsky, and P. Hoyer, *Nucl. Phys.* **B360** 67 (1991); Nucl. Phys. **B383** 643 (1992); R. Vogt and S. J. Brodsky, SLAC-PUB-6468 (1994), Nucl. Phys. **B**, in press.
17. V. Papadimitriou, CDF Coll., FERMILAB-Conf-94/221-E.
18. E. Braaten and S. Fleming, Northwestern preprint, NUHEP-TH-94-26, hep-ph/9411365.
19. G. T. Bodwin, E. Braaten, and G. P. Lepage, *Phys. Rev.* **D46**, 1914 (1992).
20. S. J. Brodsky, P. Hoyer, A. H. Mueller, and W.-K. Tang, Nucl. Phys. **B369**, 519 (1992).
21. E. L. Berger and S. J. Brodsky, Phys. Rev. Lett. **42**, 940 (1979).
22. K. J. Eskola, P. Hoyer, M. Vänttinen, and R. Vogt, *Phys. Lett.* **B333**, 526 (1994).
23. J. G. Heinrich, *et al.*, *Phys. Rev.* **D44**, 1909 (1991).
24. A. Brandenburg, S. J. Brodsky, V. V. Khoze, and D. Müller, Phys. Rev. Lett. **73**, 939 (1994).

25. E. Braaten and T. C. Yuan, Phys. Rev. Lett. **71**, 1673 (1993).
26. E705: L. Antoniazzi, *et al.*, Phys. Rev. Lett. **70**, 383 (1993).
27. E672: A. Zieminski, *et al.*, *Proceedings of the XXVI International Conference on High Energy Physics*, Dallas, Texas, 1992, AIP Conference Proceedings No. 272, Ed. by J. R. Sanford, p. 1062.
28. M. Vänttinen, P. Hoyer, S. J. Brodsky, and W. K. Tang, SLAC-PUB-6637 (1994).
29. G. A. Schuler, CERN-TH.7170/94 (February 1994).
30. NA3: J. Badier, *et al.*, Z. Phys. **C20**, 101 (1983).
31. C. Biino, *et al.*, Phys. Rev. Lett. **58**, 2523 (1987).
32. E537: C. Akerlof, *et al.*, Phys. Rev. **D48**, 5067 (1993).
33. B. L. Ioffe, Phys. Rev. Lett. **39**, 1589 (1977).
34. E. L. Berger and D. Jones, Phys. Rev. **D23**, 1521 (1981).
35. E. N. Argyres and C. S. Lam, Phys. Rev. **D26**, 114 (1982).
36. E. N. Argyres and C. S. Lam, Nucl. Phys. **B234**, 26 (1984).
37. R. Baier and R. Rückl, Nucl. Phys. **B201**, 1 (1982).
38. R. Baier and R. Rückl, Nucl. Phys. **B218**, 289 (1983).
39. J. G. Körner, J. Cleymans, M. Kuroda, and G. J. Gounaris, Nucl. Phys. **B204**, 6 (1982).
40. E772: D. M. Alde, *et al.*, Phys. Rev. Lett. **64**, 2479 (1990).
41. E772: D. M. Alde, *et al.*, Phys. Rev. Lett. **66**, 133 (1991).
42. P. Hoyer, M. Vänttinen, and U. Sukhatme, Phys. Lett. **B246**, 217 (1990).
43. S. J. Brodsky, J. C. Collins, S. D. Ellis, J. F. Gunion, and A. H. Mueller, in *Proceedings of the Summer Study on the Design and Utilization of the Superconducting Super Collider,* Snowmass, CO, 1984, edited by R. Donaldson and J. Morfin (Division of Particles and Fields of the American Physical Society, New York, 1985).
44. S. J. Brodsky, P. Hoyer, C. Peterson and N. Sakai, *Phys. Lett.* **B93**,451 (1980); S. J. Brodsky, C. Peterson and N. Sakai, *Phys. Rev.* **D23**, 2745 (1981).
45. J. J. Aubert *et al.*, *Phys. Lett.* **110B**, 73 (1982). E. Hoffmann and R. Moore, *Z. Phys.* **C20**, *71 (1983)*.
46. *M. Adamovich et al.*, *Phys. Lett.* **B305**, 402 (1993). G. A. Alves *et al.*, *Phys. Rev. Lett.* **72**, 812 (1994).
47. J. Badier *et al.*, *Z. Phys.* **C20**, 101 (1983).
48. C. Biino *et. al.*, *Phys. Rev. Lett.* **58**, 2523 (1987). M. Vänttinen, P. Hoyer, S. J. Brodsky, and W.-K. Tang, SLAC-PUB-6637 (1994).
49. J. Badier *et al.*, *Phys. Lett.* **114B**, 457 (1982).
50. J. Badier *et al.*, *Phys. Lett.* **158B**, 85 (1985).

51. V. Barger, F. Halzen, and W.Y. Keung, *Phys. Lett.* **119B**, 453 (1982).
52. R. E. Ecclestone and D.M. Scott, *Phys. Lett.* **120B**, 237 (1983).
53. B. Humpert and P. Mery, *Phys. Lett.* **124B**, 265 (1983).
54. V. G. Kartvelishvili and Sh. M. Ésakiya, *Sov. J. Nucl. Phys.* **38**(3), 430 (1983) [Yad. Fiz. **38**, 722 (1983)].
55. B.-A. Li and K.-F. Liu, *Phys. Rev.* **D29**, 426 (1984).
56. R. Vogt and S. J. Brodsky SLAC-PUB-95-6753, (1995).
57. S. D. Drell and A. C. Hearn, *Phys. Rev. Lett.* **16**,908 (1966); S. Gerasimov, Yad. Fiz. **2**,598 (1965) [*Sov. J. Nucl. Phys.* **2**, 430 (1966)]; L. I. Lapidus and Chou Kuang-Chao, *J. Exptl. Theoretical Physics* **41**, 1545 (1961) [*Sov. Phys. JETP* **14**, 1102 (1962)]; M. Hosada and K. Yamamoto, *Prog. Theor. Phys.* **36**, 426 (1966). For a recent review of the empirical tests of the DHG sum rule see B. L. Ioffe, preprint ITEP-61 (1994); D. Drechsel, University of Mainz preprint, 1994.
58. S. J. Brodsky, S. D. Drell, *Phys. Rev.* **D22**,2236 (1980). For a recent review of the empirical bounds on the radii and the anomalous moments of leptons and quarks, see G. Köpp, D. Schaile, M. Spira, and P. M. Zerwas, DESY preprint DESY-94-148.
59. S. J. Brodsky and J. R. Hiller, *Phys. Rev.* **D46**,2141 (1992).
60. S. J. Brodsky and J. R. Primack, *Annals Phys.* **52** 315 (1969); *Phys. Rev.* **174**, (1968) 2071.
61. S. J. Brodsky and F. Schlumpf, *Phys. Lett.* **B329**, 111 (1994).
62. B. Q. Ma, *J. Phys.* **G17**, L53 (1991); Bo-Qiang Ma and Qi-Ren Zhang, *Z. Phys. C* **58**, (1993) 479.
63. S. Weinberg, Lectures on elementary particles and field theory, Vol. 1, eds. S. Deser, M. Grisaru and H. Pendleton (MIT Press, 1970, Cambridge, USA).
64. S. Ferrara, M. Porrati, and V. L. Telegdi, *Phys. Rev.* **D46**, 3529 (1992).
65. V. Bargmann, L. Michel, and V. L. Telegdi, *Phys. Rev. Lett.* **2**, 435 (1959).
66. R. W. Brown, K. L. Kowalski and S. J. Brodsky, *Phys. Rev.* **D28**, 624 (1983); S. J. Brodsky and R. W. Brown, *Phys. Rev. Lett.* **49**, 966 (1982).
67. F. E. Low, *Phys. Rev.* **96**,1428 (1954); *Phys. Rev.* **110**, 974 (1958). M. Gell-Mann and M. L. Goldberger, *Phys. Rev.* **96**, 1433 (1954).
68. K. O. Mikaelian, M. A. Samuel and D. Sahdev, *Phys. Rev. Lett.* **43**, 746 (1979).
69. G. Altarelli, N. Cabibbo and L. Maiani, *Phys. Lett.* **40B**, 415 (1972).
70. S. J. Brodsky and I. Schmidt, SLAC-PUB 95-6761 (1995).
71. I. F. Ginzburg, G. L. Kotkin, S. L. Panfil and V. G. Serbo, *Nucl. Phys.* **B228**,285 (1983). See also S. Y. Choi and F. Schrempp, *Phys. Lett.* **B272**,149 (1991); M. Raidal, Helsinki preprint HU-SEFT R 1994-16.
72. S. J. Brodsky, T. Rizzo and I. Schmidt, in preparation.

73. S. J. Brodsky and H. J. Lu, SLAC-PUB-6481 (1994).
74. S. J. Brodsky, G. P. Lepage and P. B. Mackenzie, *Phys. Rev.* **D28,** 228 (1983).
75. R. J. Crewther, *Phys. Rev. Lett.* **28,** 1421 (1972).
76. D. J. Broadhurst and A. L. Kataev, *Phys. Lett.* **B315,** 179 (1993).
77. E.C.G. Stückelberg and A. Peterman, *Helv. Phys. Acta* **26,** 499 (1953), A. Peterman, *Phys. Rept.* **53C,** 157 (1979).
78. M. Neubert, CERN-TH-7524-94, (1995).
79. P. Ball, M. Beneke, and V. M. Braun, CERN-TH-95-26, (1995).
80. G. Ingelman and J. Rathsman, preprint TSL-ISV-94-0096.
81. G. Kramer and B. Lampe, *Zeit. Phys.* **A339,** 189 (1991).
82. B. H. Smith and M. B. Voloshin, preprint TPI-MINN-94-16-T.
83. A. Sirlin, preprint NYU-TH-94-08-01.